# Parabolic Scaling in Overdoped Cuprate Films

Yong Tao[†]

College of Economics and Management, Southwest University, Chongqing, China
Department of Management, Technology, and Economics, ETH Zurich, Switzerland

**Abstract:** It was recently reported that, in the highly overdoped side of single-crystal $La_{2-x}Sr_xCuO_4$ films, the transition temperature $T_c$ and zero-temperature superfluid phase stiffness $\rho_s(0)$ will obey a parabolic scaling $T_c = \gamma \cdot \sqrt{\rho_s(0)}$. Parabolic scaling indicates a quantum phase transition from a superconductor to a normal metal, for which there has been scant understanding [Nature **536**, 309-311 (2016)]. The current study shows that, using the quantum critical model for zero-temperature Cooper pairs [EPL **118,** 57007 (2017)], parabolic scaling can be exactly derived, where $\gamma = \gamma(\varepsilon_F, a)$ is uniquely determined by the Fermi energy $\varepsilon_F$ and the minimal lattice constant $a$ of superconducting materials. For single-crystal $La_{2-x}Sr_xCuO_4$ films, we calculate the theoretical value of $\gamma$, which yields $4.29 \cdot K^{1/2}$ and is in accordance with an experimental measure value $(4.2 \pm 0.5) \cdot K^{1/2}$ with high accuracy. Our formula for $\gamma$ can be further tested by investigating other BCS-like materials.

---

[†] Correspondence to: taoyingyong@yahoo.com or taoy@ethz.ch

# 1. Introduction

To explore the potential origin of high-temperature superconductivity in copper oxide materials, much research has focused on the universal correlations among physical quantities controlling the superconducting mechanism. One of the earliest patterns was called Homes' law [1-2]: $T_c \propto \rho_s(0)/\sigma_{dc}$, which marks the linear scaling between the transition temperature $T_c$ and the zero-temperature superfluid phase stiffness $\rho_s(0)$, where $\sigma_{dc}$ denotes the d.c. conductivity measured at approximately $T_c$. Homes' law became well-known as a mean-field result of the dirty-limit BCS theory [2-3] and was expected to hold regardless of underdoped, optimally doped, and overdoped materials [1]. However, some scholars questioned the validity of Homes' law in highly underdoped [4-6] and overdoped materials [7]. Recently, by investigating the overdoped side of single-crystal $La_{2-x}Sr_xCuO_4$ films, Bozovic *et al.* observed that $T_c$ and $\rho_s(0)$ obeyed two-class scaling [8]:

$$\begin{cases} T_c = \alpha \cdot \rho_s(0) + T_0, & T_c \geq T_M \\ T_c = \gamma \cdot \sqrt{\rho_s(0)}, & T_c \leq T_Q \end{cases}, \qquad (1)$$

where $T_M \approx 12K$, $T_Q \approx 15K$, $\alpha = 0.37 \pm 0.02$, $T_0 = (7.0 \pm 0.1) \cdot K$, and $\gamma = (4.2 \pm 0.5) \cdot K^{1/2}$. In particular, two-class scaling equation (1) is non-smoothly linked by linear and parabolic parts; that is, there is a non-smooth kink over the $T_c$ interval $[T_M, T_Q]$ [8]. Two-class scaling equation (1) differs significantly from Homes' law; therefore, Bozovic *et al.* concluded that their experimental findings were incompatible with the mean-field description [8]. Later, some phenomenological theories [9-10] were proposed to explain two-class scaling equation (1). For example, the dirty *d*-wave BCS theory [9], a mean-field model, was proposed, which leads to smooth two-class scaling, eliminating the kink over the interval $[T_M, T_Q]$. Although all of these phenomenological theories [9-10] can produce parabolic scaling $T_c \propto \sqrt{\rho_s(0)}$ and are well fitted with experimental data by resorting to some phenomenological parameters, they cannot reproduce the experimental value of the coefficient $\gamma$ in parabolic scaling. Indeed, due to the presence of phenomenological parameters, it is difficult to experimentally distinguish which phenomenological theory is more reasonable. From the perspective of falsifiability, a rigid physical theory should produce theoretical values in accordance with the existing experimental results.

Recently, Tao proposed a quantum critical model [11-12] to show that two-class scaling equation (1) is due to two different physical mechanisms [12]: linear scaling is a mean-field behavior of the dirty-limit BCS theory, while parabolic scaling is a quantum critical behavior. The non-smooth kink over the interval $[T_M, T_Q]$ is a result of the competition between two mechanisms. Tao's theory [12] is a principle model (without phenomenological parameters) rather than a phenomenological model, and hence can be tested rigidly. The current study shows that, different from phenomenological theories, Tao's model leads to the exact parabolic scaling

$$T_c = \gamma \cdot \sqrt{\rho_s(0)}, \tag{2}$$

where the theoretical value of $\gamma$ is in accordance with the existing experimental result with high accuracy.

## 2. Quantum critical model for zero-temperature Cooper pairs

Before beginning our computation, we first outline the basic idea of Tao's model [12]. Here we adopt the natural units $\hbar = c = k_B = 1$, where $\hbar$ denotes the reduced Planck constant, $c$ is the light speed, and $k_B$ is the Boltzmann constant. Tao's model is based on the following three steps [12].

First, we expand the order parameter $\phi(T)$ around $T_c$ to obtain the free energy density of the Landau-Ginzburg equation:

$$\mathcal{L}(T, T_c) = \mathcal{L}_0 + |\nabla \phi(T)|^2 + \lambda_2(T, T_c) \cdot |\phi(T)|^2 + \lambda_4(T, T_c) \cdot |\phi(T)|^4, \tag{3}$$

where $|T - T_c| \approx 0$ and $\nabla$ denote the spatial derivative. Here $\lambda_2(T, T_c)$ and $\lambda_4(T, T_c)$ are undetermined parameters.

Second, by applying Gor'kov's Green function method into the BCS theory [13-14], we determine the coefficients $\lambda_2(T, T_c)$ and $\lambda_4(T, T_c)$ as follows [11-12]:

$$\lambda_2(T, T_c) = \frac{2m^* \cdot (T - T_c)}{\lambda T_c}, \tag{4}$$

$$\lambda_4(T, T_c) = \frac{2m^{*2}}{\lambda \cdot n_s(0)}, \tag{5}$$

with $\lambda = \frac{7\zeta(3) \cdot \varepsilon_F}{6\pi^2 T_c^2}$, where $n_s(0)$ denotes the zero-temperature superfluid density, $\zeta(x)$ is the Riemann zeta function, $\varepsilon_F$ is the Fermi energy of superconducting materials, and $m^*$ is the effective mass of a Cooper pair.

Third, $T_c$ is assumed to be sufficiently low so that $\mathcal{L}(T, T_c)$ is valid at $T = 0$. Since quantum fluctuations around $T = 0$ cannot be omitted, equation (3) at $T = 0$ yields [12]:

$$\mathcal{L}(0, T_c) = |\partial_\tau \phi(0)|^2 + |\nabla \phi(0)|^2 + \lambda_2(0, T_c) \cdot |\phi(0)|^2$$
$$+ \lambda_4(0, T_c) \cdot |\phi(0)|^4, \tag{6}$$

where $\partial_\tau$ denotes the imaginary-time derivative, and the imaginary time $\tau \in \left[0, \frac{1}{T}\right]$ with $T = 0$. Different from thermal critical behaviors ($T > 0$), the case of $T = 0$ is referred to as the quantum critical behavior, which requires that the order parameter $\phi(0)$ is a function of space $\boldsymbol{q}$ and imaginary time $\tau$ [15]; that is, $\phi(0) = \phi(\boldsymbol{q}, \tau)$. Based on the relativistic perspective (such as the Klein-Gordon equation), the coefficient of $|\partial_\tau \phi(0)|^2$ has been set to be same as that of $|\nabla \phi(0)|^2$ [12]. It is easy to confirm that $|\phi(0)|^2$ should denote the zero-temperature superfluid phase stiffness. To this end, equation (6) is varied to obtain the field equation of zero-temperature Cooper pairs:

$$\partial_\tau^2 \phi(0) + \nabla^2 \phi(0) - \lambda_2(0, T_c) \phi(0) - 2\lambda_4(0, T_c) \cdot |\phi(0)|^2 \phi(0) = 0.$$

For homogenous superconductors, the field equation yields $|\phi(0)|^2 = -\lambda_2(0, T_c)/2\lambda_4(0, T_c) = \frac{n_s(0)}{2m^*}$, where equations (4) and (5) have been used. Since $\frac{n_s(0)}{2m^*}$ denotes the zero-temperature superfluid phase stiffness of homogenous materials [8], $|\phi(0)|^2$ indeed denotes the zero-temperature superfluid phase stiffness.

Equation (6) is the starting point of Tao's model [12]. Its validity is justified by experimental investigation result (1). The main purpose of Tao's model is to thoroughly investigate the behaviors of $\lambda_2(0, T_c)$ and $\lambda_4(0, T_c)$ when $T_c \to 0$. From equations (4) and (5), $\lambda_2(0, T_c)$ and $\lambda_4(0, T_c)$ are derived using the BCS theory, which assumes that quantum fluctuations on all size scales are averaged out. Based on such an assumption of the mean-field, the zero-temperature superfluid density $n_s(0)$ in $\lambda_4(0, T_c)$ is equal to the total number density of electrons in the normal state [13-14] and hence can be regarded as a constant. This is the standard explanation of the BCS theory. However, Tao argued that [12], when the transition point $T_c \to 0$, quantum fluctuations around zero temperature will be amplified so that the mean-field approximation may break down. This implies that the behaviors of $\lambda_2(0, T_c)$ and $\lambda_4(0, T_c)$ at $T_c \to 0$ should be affected by quantum fluctuations, and somewhat differ from the mean-field description. For example, $n_s(0)$ may change when $T_c \to 0$.

To deal with quantum fluctuations, we need to assess the quantum partition function of equation (6) at $T = 0$ [12]:

$$Z(0, T_c) = \int [\mathcal{D}\phi^*(0)]_\Lambda \int [\mathcal{D}\phi(0)]_\Lambda \, e^{-\int d\tau \int d^D \boldsymbol{q} \cdot \mathcal{L}(0, T_c)}, \tag{7}$$

where $\Lambda$ denotes the momentum cut-off, which implies that quantum fluctuations with length scales less than $\frac{1}{\Lambda}$ can be omitted. Here $D$ denotes the dimension of superconducting materials. Equation (7) with $T_c \to 0$ is Tao's model [12].

## 3. Parabolic scaling in $La_{2-x}Sr_xCuO_4$ films

This section shows how Tao's model leads to the exact parabolic scaling equation (2). To deal with quantum fluctuations, Tao [12] introduced the renormalization procedure into equation (7). Wilson [16-18] proposed that any quantum field theory should be defined fundamentally with a cut-off $\Lambda$ that has some physical significance. Here we adopt Wilson's proposal. For crystal materials, a rigid renormalization theory can be defined on a cubic lattice of a lattice unit [16-17]:

$$a = \frac{1}{\Lambda}, \tag{8}$$

where $a$ denotes the minimal lattice constant of superconducting materials. Obviously, single-crystal $La_{2-x}Sr_xCuO_4$ films are crystal materials [8]. Equation (8) can be regarded as a result of the Bohr-Sommerfeld quantization condition $\oint p \cdot dq = nh$ with $n = 1$. From this meaning, the minimal lattice constant $a$ plays the role of a "Bohr radius"; that is, $a$ determines the stability of crystal structures as the Bohr radius has for atoms. Therefore, the physical meaning of equation (8) is that quantum fluctuations with wavelengths less than $2\pi a$ can be averaged out [18]. Weinberg also pointed out that, in solid-state physics, there really is a cut-off, the lattice spacing $a$, which one must take seriously in dealing with phenomena at similar length scales[1].

Applying the renormalization group approach to equation (7), Tao proved that [12], when $T_c \to 0$, $\lambda_4(0, T_c)$ yields a fixed point depending on the momentum cut-off $\Lambda$, that is

$$\lambda_4(0, T_c \to 0) = \text{const} \cdot \Lambda^{3-D}, \tag{9}$$

which, up to the one-loop correction, leads to parabolic scaling [12]:

$$T_c = \sqrt{(3-D) \cdot \Lambda^{3-D} \cdot \frac{7(2\pi)^D \Gamma\left(\frac{D}{2}\right) \zeta(3) \cdot \varepsilon_F}{30(\pi)^{\frac{D}{2}+2} m^*}} \cdot \sqrt{\rho_s(0)}, \tag{10}$$

where $\rho_s(0) = \frac{n_s(0)}{2m^*}$.

At $T = 0$, we assume:

---

[1] Private correspondence from Professor Steven Weinberg.

$$m^* = 2 \cdot m_e, \tag{11}$$

where $m_e$ denotes the rest mass of an electron.

Since Bozovic *et al.* investigated single-crystal $La_{2-x}Sr_xCuO_4$ films [8], we use $D = 2$. Comparing equations (2) and (10), substituting $D = 2$, equations (8) and (11) into equation (10) obtains:

$$\gamma(\varepsilon_F, a) = \sqrt{\frac{7 \cdot \zeta(3) \cdot \varepsilon_F}{15 \cdot \pi \cdot a \cdot m_e}}. \tag{12}$$

Using equation (12), the theoretical value of $\gamma$ can be calculated using the experimental values of $\varepsilon_F$ and $a$. Equation (12) is the main theoretical result of this paper. It universally holds for any quasi-two-dimensional BCS-like superconductors and can be thoroughly tested. Here we show that, for single-crystal $La_{2-x}Sr_xCuO_4$ films, it provides the theoretical value in accordance with the experimental result.

Substituting $\hbar = c = k_B = 1$ into $\hbar \approx 6.58 \times 10^{-16} \cdot eV \cdot s$, $c \approx 2.99 \times 10^8 \cdot m \cdot s^{-1}$, and $k_B \approx 8.62 \times 10^{-5} \cdot eV \cdot K^{-1}$ obtains:

$$1 \cdot s \approx \frac{1}{6.58} \times 10^{16} \cdot eV^{-1}, \tag{13}$$

$$1 \cdot m \approx \frac{1}{2.99} \times 10^{-8} \cdot s, \tag{14}$$

$$1 \cdot eV \approx \frac{1}{8.62} \times 10^5 \cdot K. \tag{15}$$

The reference [8] provides the minimal lattice constant of $La_{2-x}Sr_xCuO_4$ as follows:

$$a \approx 3.8 \times 10^{-10} \cdot m. \tag{16}$$

Moreover:

$$m_e \approx 0.51 \times 10^6 \cdot eV \tag{17}$$

$$\zeta(3) \approx 1.2 \tag{18}$$

Bozovic *et al.* investigated single-crystal $La_{2-x}Sr_xCuO_4$ films around $x = 0.25$ [8]. Regarding $La_{2-x}Sr_xCuO_4$, Kamimura and Ushio's numerical calculation showed that the Fermi energy at $x \approx 0.2$ yields [19]:

$$\varepsilon_F \approx 8.75 \cdot eV. \tag{19}$$

Substituting the data from equations (13)-(19) into equation (12) obtains the theoretical value of $\gamma$:

$$\gamma_{theory} \approx 4.29 \cdot K^{1/2}. \tag{20}$$

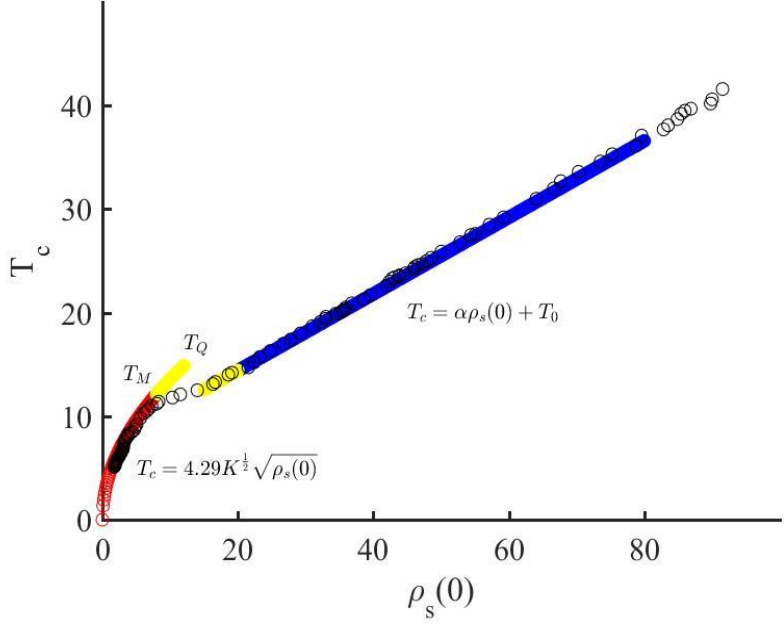

**Figure 1:** The experimental data from [8] are plotted as black circles, which belong to the $T_c$ interval $[5.1K, 41.6K]$. The parabolic scaling (red line) $T_c = 4.29\, K^{1/2} \cdot \sqrt{\rho_s(0)}$ perfectly fits the experimental data in $[5.1K, T_M]$, while the linear scaling (blue line) perfectly fits the experimental data in $[T_Q, 41.6K]$. Both parabolic and linear scaling (yellow lines) fail to fit the experimental data in $[T_M, T_Q]$, where $T_M \approx 12K$ and $T_Q \approx 15K$.

The experimental value measured by Bozovic *et al.* is as follows [8]:

$$\gamma_{experiment} = (4.2 \pm 0.5) \cdot K^{1/2}. \tag{21}$$

By equations (20) and (21), the theoretical value of $\gamma$ is in accordance with the experimental value with surprisingly high accuracy. For a visual comparison, we employed theoretical parabolic scaling, $T_c = 4.29\, K^{1/2} \cdot \sqrt{\rho_s(0)}$, to fit the experimental data in Figure 1, where the theoretical formula (red line) perfectly fits the experimental data in the $T_c$ interval $[5.1K, T_M]$. The theoretical and experimental values of $\gamma$ are listed in Table 1. It is quite impressive that the theoretical result given by equation (10) at $D = 2$ is in accordance with the experimental value with such high accuracy. This result thoroughly proves the validity of equation (7) in exactly describing the quantum critical behaviors of zero-temperature Cooper pairs. Therefore, using equations (6) and (10), the field equation of zero-temperature Cooper pairs at $T_c \to 0$ should exactly follow:

$$\partial_\tau^2 \phi(0) + \nabla^2 \phi(0) - \lambda_2(0, T_c) \cdot \phi(0) + \frac{\lambda_2(0,T_c)}{\rho_s(0)} \cdot |\phi(0)|^2 \cdot \phi(0) = 0. \tag{22}$$

Finally, we investigate the hidden physical meaning of equation (10). Since $\rho_s(0)$ is proportional to $n_s(0)$, equation (10) implies that $n_s(0)$ will decrease as $T_c$ decreases. This significantly differs from the standard explanation of the BCS theory, where $n_s(0)$ is equal to the total number density of electrons in the normal state. In fact, the BCS theory is a mean-field theory that assumes that quantum fluctuations on all size scales are averaged out. When the transition point $T_c \gg 0$, we believe that the mean-field description is a good approximation. However, once the transition point $T_c \to 0$, quantum fluctuations around zero temperature will be amplified. Then the mean-field approximation may break down. To deal with quantum fluctuations, we must introduce the renormalization procedure into equation (6). This means that $\lambda_4(0, T_c)$ will receive the contributions from quantum fluctuations, and thus changes with the momentum cut-off $\Lambda$. As a result, the zero-temperature superfluid density $n_s(0)$ is no longer a constant. This implies that, when $T_c \to 0$, quantum fluctuations break the Cooper pairs. Indeed, Bozovic *et al.* found that, with increased doping ($T_c \to 0$), $La_{2-x}Sr_xCuO_4$ becomes more metallic, and increased doping induces a quantum phase transition from a superconductor to a normal metal [20-21]. Here we show that such a phase transition can be well described by equation (7). To see this, we focus on the free energy density (6) that takes the minimum at

$$|\phi(0)|_{vac} = \sqrt{\frac{n_s(0)}{2m^*}} = \sqrt{\rho_s(0)}, \tag{23}$$

which using equation (10) yields:

$$|\phi(0)|_{vac} = \frac{1}{\gamma} \cdot T_c. \tag{24}$$

For $D = 2$, equation (24) clearly indicates a phase transition point $T_c = 0$.

## 4. Conclusion

In the highly overdoped side, with increased doping, $La_{2-x}Sr_xCuO_4$ becomes more metallic, and increased doping induces a quantum phase transition from a superconductor to a normal metal [20]. Quantitatively, it can be described as a parabolic scaling between the zero-temperature superfluid phase stiffness $\rho_s(0)$ and the transition temperature $T_c$. This paper showed that the parabolic scaling can be exactly derived using the quantum critical model (7), which describes the behaviors of zero-temperature Cooper pairs at $T_c \to 0$. In particular, for two-dimensional BCS-like systems, the order parameter $\phi(0)$ in quantum critical model (7) clearly indicates a phase transition point $T_c = 0$. Based on this model, we calculated the theoretical value of the coefficient $\gamma$ in the parabolic scaling, which agrees with the experimental measure value with surprisingly high accuracy. Our results imply that, for quasi-two-dimensional BCS-like superconductors, when the transition temperature $T_c$ approaches zero, quantum fluctuations should play a central role in breaking Cooper pairs. In this sense, the phase transition from a superconductor to a normal metal in the overdoped side of $La_{2-x}Sr_xCuO_4$ is induced by quantum fluctuations.

## Acknowledgments

This work was supported by the Fundamental Research Funds for the Central Universities (Grant No. SWU1409444 and Grant No. SWU1809020), the National Natural Science Foundation of China (Grant No.71773099), and the State Scholarship Fund granted by the China Scholarship Council. The author acknowledges Professor Steven Weinberg for confirming the validity of the lattice spacing as the (ultraviolet) cut-off in solid-state physics.

**Table 1.** Comparison of the theoretical result with the experimental measure value [8] for the coefficient in parabolic scaling equation (2)

| Coefficient | Experimental value | Theoretical result |
| --- | --- | --- |
| $\gamma$ | $(4.2 \pm 0.5) \cdot K^{1/2}$ | $4.29 \cdot K^{1/2}$ |